\documentclass[aps,pre,showpacs,twocolumn]{revtex4-1}
\usepackage{graphicx}
\begin{document}

\title{Efficient Parallelization of Short-Range Molecular Dynamics Simulations on Many-Core Systems}

\author{R. Meyer}
\email{rmeyer@cs.laurentian.ca}

\affiliation{Department of Mathematics and Computer Science and Department of Physics,
 Laurentian University, 935 Ramsey Lake Road, Sudbury (Ontario) P3E 2C6, Canada}

%
%
\begin{abstract}
This article introduces a highly parallel algorithm for molecular dynamics simulations with short-range forces on single node multi- and many-core systems. The algorithm is designed to achieve high parallel speedups for strongly inhomogeneous systems like nanodevices or nanostructured materials. 
In the proposed scheme the calculation of the forces and the generation of neighbor lists are divided into small tasks. The tasks are then executed by a thread pool according to a dependent task schedule. This schedule is constructed in such a way that a particle is never accessed by two threads at the same time.
Benchmark simulations on a typical 12 core machine show that the described algorithm achieves excellent parallel efficiencies above 80\,\% for different kinds of systems and all numbers of cores. For inhomogeneous systems the speedups are strongly superior to those obtained with spatial decomposition. Further benchmarks were performed on an Intel Xeon Phi coprocessor. These simulations demonstrate that the algorithm scales well to large numbers of cores. 
\end{abstract}

\date{November 18, 2013}
\maketitle

%
%
\section{Introduction}\label{Intro}
Molecular dynamics (MD) simulation is one of the most important  numerical methods in computational physics, chemistry, biology and materials science \cite{AllenTildesley, FrenkelSmit,Rapaport}. It is a very versatile method that allows particle based simulations of a wide variety of systems provided that a suitable model for the interactions between the particles exists.  Advances of computer technology  have dramatically increased the number of particles that can be handled by MD simulations. When used in conjunction with a short-range interaction model, i.e.\ a model where particle interactions become zero if the distance between the particles exceeds a cutoff distance $r_\mathrm{cut}$,  simulations containing tens or hundreds of millions of particles can be performed without too many difficulties.

In recent years, a strong increase of computational power has been achieved through the introduction of multicore processors that integrate multiple CPU cores on a single chip. Currently the trend goes from multi-core to many-core systems which integrate hundreds or thousands of processors. An early example of this technology is the Intel Xeon Phi coprocessor \cite{JeffersReinders}. Such massively parallel systems both enable and require new approaches to the development of parallel programs.

In this article a task-based approach to the parallelization of MD simulations is described. The \textit{cell task} algorithm overcomes some limitations of previous approaches and gives excellent parallel performance in simulations of large-scale systems. In its current form the algorithm is designed for single node systems with approximately uniform memory access times. Applicable systems range from dual and quad core desktop computers over  typical HPC nodes or server systems (8 - 48 cores) to the Intel Xeon Phi (60 cores, 4 hardware threads per core). An extension of the algorithm into a hybrid scheme that uses message passing between nodes is straightforward. The algorithm is not designed for shared memory systems with strongly non-uniform memory access times (so-called virtual shared memory machines) or GPU computing. The implementation of MD simulations on GPUs has recently been discussed in Ref.~\cite{Anderson:08a}.

The motivation for the development of a new MD algorithm is twofold. On the one hand, the cell task method is an attempt to explore new ways to adapt to the changes in computing technology and to get the most out of modern hardware. On the other hand, the applications of the MD method have changed over time. It is now possible to simulate complete nanodevices or nanostructured materials that combine different materials on length scales of several nanometers.  As discussed below, the inhomogeneity of such systems reduces the efficiency of previous methods. The proposed algorithm has been designed specifically to provide an efficient parallelization for such systems.

A necessary ingredient for large-scale MD simulations is parallel computing. Three principal strategies for the parallelization of MD simulations are space (or domain) decomposition, particle decomposition, and force decomposition \cite{Berendsen:95a,Plimpton:95a}. On shared memory architectures one can also employ thread-based approaches using, for example, OpenMP \cite{OpenMP,OpenMPBook}.   

The particle decomposition method and the force decomposition method are both based on a static decomposition of the system's force matrix. Particle decomposition assigns complete rows of the force matrix to a processor while force decomposition uses a block decomposition of this matrix. As discussed by Plimpton~\cite{Plimpton:95a}, both methods are not ideal for simulations of large systems with short-range forces due to communication overhead. In addition to this, the force decomposition method is subject to load balancing issues if the force matrix is not uniformly sparse.

In the spatial decomposition approach (see, e.g., Ref. \cite{Plimpton:95a}), the simulation cell is divided into as many domains as there are processors. Each processor is then responsible for the calculations of forces on particles in one domain. When a particle crosses the border between two domains, it is reassigned to the processor of the new domain. Spatial decomposition achieves very good parallel speedups under two conditions: the domains must be large enough so that most of the interactions happen between particles on the same processor, and the particle density system must be sufficiently homogeneous in order to achieve comparable computational loads on the processors.  
Strongly inhomogeneous systems do not satisfy the latter condition, and the efficiency of spatial decomposition for these systems is therefore reduced by load imbalances.

Recently, a number of variations of the spatial decomposition approach have been developed \cite{Snir:04a,Shaw:05a,Bowers:05a,Bowers:06a}. These neutral territory methods have the potential to outperform  traditional spatial decomposition for high levels of parallelism.  These methods are, however, not well suited to handle strongly inhomogeneous systems.  

While it is fairly straightforward to parallelize MD programs using OpenMP, this approach rarely leads to  satisfactory speedups. The reason for this are uncoordinated accesses to the particle data by the threads. In order to avoid race conditions particle updates must be protected by synchronization constructs or Newton's third law cannot be exploited. The performance is further degraded since accesses to the same particle by different threads may result  in frequent transfers  of cache lines between the CPU's. 

Task-based programming is a modern approach to parallel computing. This technique uses high-level abstractions to subdivide the problem into small work units (tasks) without regard for details of the task execution on the hardware. These techniques are generally most efficient if the number of tasks that can be executed concurrently is much larger than the number of available processors. Task-based programming and examples of  task-based programming models are discussed in Refs.~\cite{Korch:04,Planas:09,Vandierendonck:11,Rauber:12}.
 
The cell task method described in this article uses a task-based programming model. The method subdivides the calculation of the forces in a large number of small tasks. These tasks are then executed according to a dependent task schedule that avoids access conflicts between the threads. For sufficiently large systems this leads to excellent parallel speedups that are largely independent of the degree of homogeneity of the simulated system. In addition to this, the algorithm relieves the user from technical considerations like the best subdivision of the system for a given number of processors.

%
%
\section{Description of the cell task algorithm}\label{SecAlgorithm}
\subsection{Partitioning of the problem into tasks}\label{SecTask}
The primary objective of this work was the design of an efficient parallel MD algorithm that achieves high parallel speedups for large and strongly inhomogeneous systems. Secondary goals were  consistent speedups for all numbers of processors and to make the method robust against external perturbations that might temporarily delay computations on individual processors. 

The problems of spatial decomposition and simple thread-based approaches can both be traced to the kind of geometric information used by the methods. Spatial decomposition fails to achieve a good load balance if the size of the processor domains exceeds the characteristic length scale of inhomogeneities in the system. In the simple thread-based  approach on the other hand the efficiency is reduced by the need for synchronization which is ultimately due to the fact that no geometric information is used at all. These observations lead to the idea to use geometric information on a small length scale for the parallelization.

In order to avoid synchronization constructs during the calculation of forces on the particles, one must ensure that two threads will never update the same particle at the same time. This means that threads must keep a distance of $2 r_\mathrm{cut}$ between the particles on which they work. Fortunately, the necessary information to ensure this condition is readily available in many cases or it can be obtained easily. Most general purpose MD codes use the so-called linked-cell technique \cite{AllenTildesley,Rapaport:91a} to facilitate the construction of neighbor lists. In the linked-cell method, the simulation box is subdivided into a grid of small cells whose width is larger than $r_\mathrm{cut}$ and for each cell a linked list of the particles in the cell is constructed. A particle can then interact only with particles in the same cell or one of its neighbors.

\begin{figure}[b]
\includegraphics[width=7.2cm]{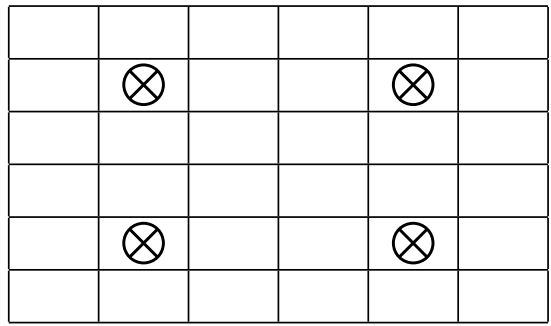}
\caption{Example for a set of non-overlapping tasks on a two-dimensional cell grid. Tasks are indicated by $\otimes$ symbols in the respective grid cells.}\label{FigWave}
\end{figure}
The proposed parallelization method reuses the data structures from the linked cell technique to partition the problem of the force calculation into tasks. A task in this method consists of the calculation of the forces on all particles in one grid cell. Since particles in one cell do only interact with particles in neighboring cells, two tasks can be executed concurrently if there are at least two grid cells between them in all directions. Tasks that fulfill this condition are called \textit{non-overlapping} otherwise they are said to be \textit{overlapping}. Figure~\ref{FigWave} shows a two-dimensional example of a cell grid and a set of tasks that can be executed in parallel since they are non-overlapping.

The essence of the proposed algorithm is the following: 
\begin{enumerate}
\item The particles are binned into grid cells (as part of the linked cell technique).
\item The force calculation is broken into cell tasks. A cell tasks consists of the calculation of forces on all particles in one grid cell.
\item The tasks are executed in parallel by a thread pool  according to a dependent schedule that ensures that only non-overlapping tasks are executed simultaneously.
\end{enumerate} 

The algorithm does not assume that the execution of the tasks requires similar amounts of computing time. Even in a homogeneous system this might not be the case and in inhomogeneous systems a  substantial amount of grid cells might in fact be empty. Notwithstanding, a good load balance is achieved if the tasks are dynamically scheduled rather then statically. Dynamic scheduling assigns the tasks to the threads as these finish previous tasks. This keeps all threads busy and averages  the imbalances. In addition to this, dynamic scheduling limits the impact of external disturbances such as unrelated processes running on one of the cores or differences in memory access times. If a processor is slowed down by external factors, other processors will take on a larger number of tasks, thereby minimizing the delay. With static scheduling a delay of one processor leaves the rest of the system idling. 

Another positive side effect of the cell task algorithm is an improved memory access pattern. Particles in the same grid cell have similar sets of neighbors. The calculation of the forces on particles in the same cell within the same task therefore leads to high cache reuse without a special ordering of the particles.

After the force calculation, the second most time-consuming part of an MD simulation is usually the generation of the neighbor lists. In a program that uses the linked cell method the neighbor lists are naturally  generated for all particles in a grid cell at once. 
For this reason it is straightforward to apply the cell task method to the generation of the neighbor lists. The only difference to the force calculation is the kind of work performed by a cell task. During the neighbor list construction a cell task constructs the neighbor lists of all particles in the task's cell. 

An important difference between the force calculation and the generation of neighbor lists is the fact that the neighbor list generation never performs simultaneous particle updates. For this reason, the tasks generating the neighbor lists could all be scheduled simultaneously.  On the other hand, the overhead of the dependent schedule is very low as long as the system is large enough to keep all processors busy. This is since most of the overhead is incurred during the construction of the schedule, which is done only when the cell grid changes. 

While there are no simultaneous particle updates, simultaneous updates of the neighbor list data structure can be a problem. One way to solve this is that each task maintains a separate data structure for the neighbor lists of its particles.

Other typical parts of MD simulations like the integration of the equation of motion  are usually implemented with simple loops over the particles that access only one particle at a time. These loops can be parallelized using simple threading techniques since there are no synchronization concerns.

%
%
\subsection{Task scheduling: The wave method}\label{SecWave}
A central problem of the proposed algorithm is the generation and execution of a dependent task schedule that guarantees that only non-overlapping tasks run concurrently. This section presents two algorithms for this problem named the \textit{forward} and the \textit{backward wave method}. 

The wave methods group the cell tasks into an ordered set of so-called \textit{waves}. A wave is a set of non-overlapping tasks that can be scheduled independently. There are many ways to generate a set of waves that will cover all cells. In order to keep the method efficient, each wave should include as many cells as possible, and the number of dependencies between tasks should be low. The algorithm used in this work is described in the Appendix. 

The grouping of the tasks into waves establishes a partial ordering of the tasks. A simple scheduling method is to allow tasks in the same wave to run concurrently but to require that a wave must be completely finished before the next wave can start. The problem of this algorithm is that it imposes a barrier after each wave which reduces its efficiency. Note, however, that this scheme can easily be implemented in an environment that does not allow for dependent task scheduling.

The wave methods use the ordering imposed by the waves to order the access to the grid cells by the tasks. Each grid cell is accessed by at most 27 tasks (less  for cells at the simulation box boundary). The wave methods allow a task $T$ to execute as soon as all tasks from previous waves that access one of $T$'s cells are completed. A further simplification is possible since the preceding tasks depend on each other. For each of its cells $T$  only has to wait for the last task that accesses this cell before $T$ (since the last task has already waited for all previous tasks accessing this cell). The set of last tasks accessing any of $T$'s grid cells are called the direct predecessors of $T$. Similarly the direct successors of $T$ are those tasks which have $T$ as a direct predecessor.

The scheduling algorithm of both wave methods uses an acyclic graph algorithm similar to the one described in the documentation of Intel's Threading Building Blocks Library \cite{TBB}.  Each task has a data structure that stores its grid cell, the number of its direct predecessors, a list of its direct successors and a reference counter. The reference counter is initially set to the number of direct predecessors and all tasks that have no direct predecessors are added to the thread pools list of tasks that are ready for execution. The execution of a a task involves the following steps:
\begin{enumerate}
\item The work associated with the task is carried out (force calculation, neighbor list generation, etc.)
\item The reference counter of all direct successors is decremented atomically.
\item Successor tasks whose reference counter becomes zero are added to the ready list. 
\end{enumerate}

The purpose of the wave algorithms is to generate for each task its number of direct predecessors and the list of its successors. The forward wave method maintains a three-dimensional array $P[nx][ny][nz]$ that stores for each grid cell the index of the last task that has accessed the cell. Initially this array is initialized with a special marker (-1) that indicates that the cell has not been accessed. The algorithm loops over the waves in ascending order starting with the first wave. For each task in a wave the following steps are performed:
\begin{enumerate}
\item For all cells accessed by the task, copy the corresponding element of $P$ into a list.
\item Eliminate duplicates and the special marker from the list. The result is the task's set of direct predecessors.
\item Store the number of predecessors in the task structure.
\item For each direct predecessor, add the task to its successor list.
\item Initialize an empty successor list
\item Overwrite elements of $P$ used in step 1 with the task's index.
\end{enumerate} 

A drawback of the forward wave algorithm is the fact that when a task is created by the algorithm, only the predecessors are known. Since the tasks needs to store its list of successors rather than its predecessors it would simplify the task generation if the list of successors were known at the time when a task is created. This is achieved by the backward wave method. The backward method works like the forward method, but it replaces the array $P$ with an array $S[nx][ny][nz]$ which stores for each grid cell the index of the next task that accesses the cell. The array is again initialized with the special marker, which now means that no further task accesses the cell. The backward algorithm then creates the waves in reverse order from the last wave to the first. For each task in a wave the method performs the following steps:
\begin{enumerate}
\item For all cells accessed by the task, copy the corresponding element of $S$ into a list.
\item Eliminate duplicates and the special marker from the list. Store the result as the task's direct successor list.
\item For all direct successors: Increment the number of direct predecessors by one.
\item Initialize the task's number of direct predecessors to zero
\item Overwrite elements of S used in step 1 with the task's index.
\end{enumerate} 

%
%
\subsection{Implementation}\label{SecImpl}
In order to be able to test the cell task algorithm in practice, the method has been implemented in a general purpose parallel MD code named \textsf{mdntp} \cite{MeyerPHD}. This code has been developed for large-scale simulations using many-body potentials of the embedded-atom method type \cite{Daw:84a} or the similar Finnis-Sinclair~\cite{Finnis:84a} and tight-binding second-moment potentials~\cite{Cleri:93a}. The code already supported parallelization through spatial decomposition so that parallel speedups can be compared.

The implementation of the cell task method requires a software environment supporting task-based programming. The current implementation of the cell task method makes use of  Intel's Threading Building Blocks library (TBB) for the management of the thread pool and the tasks. TBB is a C++ library that is available under an open-source license for many platforms~\cite{TBB,ReindersTBB}.  It should be noted however that other task-based programming systems, for example those described in Refs.~\cite{Planas:09,Rauber:12} or OpenMP, could probably be used as well to implement the cell task method. 

Since \textsf{mdntp} uses the linked-cell technique for the generation of the neighbor lists, the implementation of the wave method required relatively little changes to the code since most data structures were already in place and ready to use. The majority of the code that had to be developed concerns the task scheduling. This code is concentrated in one C++ class named scheduler that provides member functions for the creation of the dependent task schedule and for the execution of the schedule.

The creation of the task schedule is performed during the creation of the neighbor lists when the particles have been binned into the cell grid. The scheduler then checks if the cell grid has changed and runs the backward wave algorithm if necessary. Since the schedule depends only on the cell grid, there is no need to recreate it after every execution.

In order to execute the task schedule, a functor object is passed to the scheduler. The scheduler then uses the execution algorithm described in Sec.~\ref{SecWave} to submit tasks to the TBB for execution. Whenever the TBB starts a task, the functor object is invoked to perform the actual work of the task. The functor thus acts as a delegate and makes it possible to perform different operations like force calculation or neighbor list generation with the same schedule. In contrast to a simple function pointer, the functor can pass additional information to the task. For example, in a replica style simulation where multiple copies of the system are simulated simultaneously the functor might carry the information on which system copy the task should operate.   

If the program is run with a single thread, a special version of the scheduler is created. This serial scheduler skips the generation of the task schedule and does not use the TBB.
 
Other changes concern mainly the force-calculation and neighbor-list generation. In the force calculation the loops over all particles had to be changed so that they run only over the particles in one grid cell. In addition to this the original C code was moved to a C++ class so that the scheduler could invoke it. Some care was required to create thread local storage for accumulated quantities like the potential energy. Finally, the code for the integration of the equations of motion was modified to use a thread-based parallelization.   

As mentioned in Sec.~\ref{SecTask}, a problem created by the parallel generation of the neighbor lists is the simultaneous update of the lists by multiple tasks. In order to avoid such conflicts, the current implementation stores the neighbor lists of each task separately. To this end, the data structure of each task contains a  C++ template std::vector$<>$. The usage of automatic memory management might incur some overhead in terms of both memory and speed. The experience with the code so far indicates, however, that the impact is small.

%
%
\subsection{Optimizations and refinements}\label{SecOpt}
%
%
\subsubsection{Empty task skipping}
For large systems with a substantial amount of empty volumes a large number of grid cells may be empty. Instead of scheduling a task for these cells, it makes sense to check for empty tasks during the generation of the task schedule and skip empty cells altogether. This may lead to a substantial reduction of the data structures and the scheduling overhead. The drawback of this optimization is that the task schedule must be regenerated every time that the linked-cell algorithm has run since  particles might have moved into a previously empty cell. Without empty-task skipping, the task schedule needs only to be regenerated if the number of grid cells changes. For systems that have a low number of empty cells it is therefore more efficient not to skip the empty cells and to save the overhead of the repeated schedule generation instead.

%
%
\subsubsection{Cell task blocking}
Another way to reduce the number of empty tasks and the number of tasks to be scheduled is to group a small block of $b_x\times b_y\times b_z$ grid cells into a single cell task. This corresponds to an enlargement of the grid cells by an integer factor for the purpose of task scheduling (the grid cells are still generated in the usual way so that the generation of neighbor lists is unaffected by this optimization).

Cell task blocking is a double edged sword. On the one hand, the reduction of the number of tasks through the blocking can reduce the scheduling overhead. In addition to this cache performance improves as long as all particles affected by a task fit into the cache memory.  

On the other hand, both advantages turn into disadvantages for larger cell blocks. The thread scheduler requires a large number of short tasks in order to obtain a good load balance, and the cache performance will decrease if the amount of memory used by the blocked task exceeds the size of the  cache. For these reasons  task blocking should be used with care.  Benchmark tests are   recommended in order to find the optimal block size for a system.

\subsubsection{Stricter definition of overlapping tasks}
In many cases, the definition of overlapping tasks given in Sec.~\ref{SecTask} is actually unnecessarily restrictive. If Newton's third law is exploited by a simulation program, neighbor lists are usually constructed in such a way that pairs of particles are accounted for only once. If particles $j$ and $k$ are within the cutoff radius, $j$ will appear in the neighbor list of $k$ or $k$ will be in the neighbor list of $j$ but not both. This is typically achieved by limiting the search for neighbors of a particle to 14 out of the 27 surrounding cells.

If the construction of neighbor lists excludes some of the surrounding cells, these cells can also be excluded from the definition of overlapping tasks as there is no risk of simultaneous particle updates in these cells. A reduction of  the overlap area of the tasks could be exploited to place more tasks in a wave, which increases the number of tasks that can run simultaneously. While it is difficult to exploit the exact arrangements of the 14 cells used for the neighbor lists, it is easily possible to reduce the size of the overlap area from 27 to $3\times 3\times 2 =  18$ cells. This would increase the number of tasks in a wave by up to 50\,\%.   
 
The current implementation of the cell task method does not employ a stricter definition of the task overlap-area for two reasons: First, there are actually potentials that access all 27 cells that surround an atom even if Newton's third law is applied. An example for this are the forces generated by the screening factor in the modified embedded-atom method \cite{Baskes:92a}. Second, this optimizations requires a tighter coupling between the neighbor list generation and the wave algorithm. The optimization will, however, most likely be included in a future version of the code.

%
%
\section{Results}\label{SecResults}
In order to test the efficiency of the proposed algorithm a series of benchmark tests involving four different configurations were carried out. The configurations are a cubic block of fcc bulk copper (1,000,188 atoms), a spherical copper nanoparticle with a diameter of 30\,nm (1,177,151 atoms), a porous system of partially sintered copper nanoparticles (1,992,220 atoms) and two sintered copper nanoparticles (57,482 atoms). These configurations are very different in terms of there homogeneity and the challenge they pose for the spatial decomposition method. 

All simulations used the tight-binding second-moment potential for copper by Cleri and Rosato \cite{Cleri:93a}. The structure of this potential is very similar to other many-body potentials of the embedded-atom method \cite{Daw:84a} or Finnis-Sinclair \cite{Finnis:84a} type. Compared to other potentials the Cleri and Rosato potentials use rather simple functions,  and the calculation of the forces with these potentials is therefore very fast. Due to the low computational complexity, memory access speed can become a limiting factor in simulations with these potentials. 

Simulations of the three larger systems (the small system) were run over a period of 100 (2500) simulation steps. All simulations regenerated the neighbor lists at every 10th step. Execution times were calculated as the average run time of five independent simulations excluding the time for initialization of the simulation and the loading or saving of the configuration.

The benchmark simulations were run on two different computers. The first was a typical dual hex-core server with Intel Xeon X5650 processors and 16 GB DDR3-1333 RAM. On this multi-core machine the original serial version of \textsf{mdntp}, the MPI based spatial decomposition version using 2 to 12 MPI ranks and the new task-based version using 1 to 12 threads were tested. The second test system was a Xeon Phi Coprocessor 5110P \cite{JeffersReinders}. Only the task-based version of the program was employed on this machine, running the code in native mode with 1 to 240 threads.

The simulations with the task-based method used task blocking with a block size of $2\times 2\times 2$ for the three large systems and a block size of $1\times 1\times 1$ for the small system. Empty task skipping was used except for the simulations of bulk copper. Simulations with the MPI version employed $3d$ and $2d$ decompositions where possible in order to obtain the most compact subvolumes. For the compact bulk system the impact of the details of the decomposition is very small. For the inhomogeneous systems the execution times certainly depend on the decomposition. The general picture, however, can be expected to remain the same. No attempt was therefore made to optimize the decomposition of these systems.

No direct comparison of the execution times of the program versions is attempted in the following. The execution times $t_\mathrm{task}(p)$ of the task-based version running with $p$ threads are without exception lower than the corresponding times $t_\mathrm{mpi}(p)$ of the MPI version running with $p$ ranks or the execution time $t_\mathrm{ref}$ of the original serial program. However the performance gains of the cell task versions are partly due to code changes when the force calculation was moved into a C++ class. Comparisons of the execution times or calculation of the speedup with respect to the fastest serial time [which is $t_\mathrm{task}(1)$] would therefore make little sense. 

In order to compare the cell task algorithm with spatial decomposition, the speedup of these algorithms with respect to the serial version that uses the same force calculation is used.  The speedups of the MPI and cell task version are thus calculated as $t_\mathrm{ref}/t_\mathrm{mpi}(p)$ and $t_\mathrm{task}(1)/t_\mathrm{task}(p)$, respectively. This definition eliminates the differences of the force calculation from the results and allows a fair comparison of the merits of both parallelization methods.
 
%
%
\subsection{Bulk copper}\label{SecBulk}
\begin{figure}
\includegraphics[width=7.2cm]{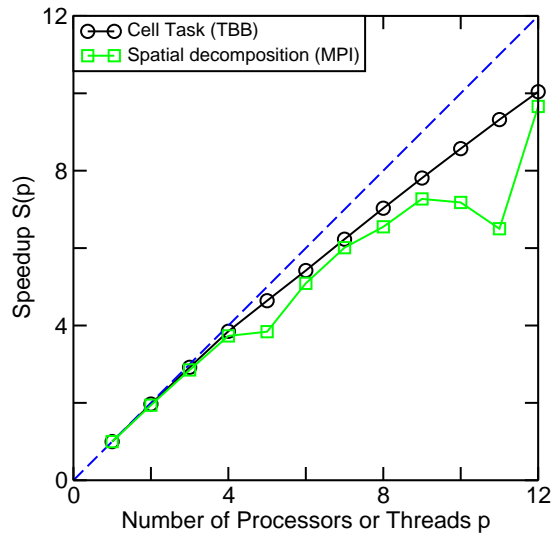}
\caption{(Color online) Parallel speedup factors in simulations of bulk copper as a function of the number of threads or processors for the cell task method (black circles) and spatial decomposition (green squares). The dashed blue line represents the ideal speedup of one per processor.}\label{FigBulk} 
\end{figure}
The crystalline bulk copper configuration is an example for a perfectly homogeneous system for which the spatial decomposition method works well. Figure~\ref{FigBulk} shows the parallel speedups obtained by the cell task method and spatial decomposition for this system on the multi-core machine. As expected, spatial decomposition yields excellent parallel efficiencies (speedup per processor or thread)  above 80\,\% for almost all numbers of processors. There are, however, dips in the curve at 5, 10, and 11 processors. The reason for this is that for these numbers the regular crystal lattice of the system cannot be divided evenly among the processors. This emphasizes another disadvantage of spatial decomposition: The efficiency of the method depends on details of the system, and not all numbers of processors work equally well.

The cell task method, on the other hand, delivers much more consistent speedup factors that increase monotonously with the number of threads. With the exception of the cases of 5, 10, and 11 processors which have already been discussed, both methods obtain similar speedups. According to the figure  the cell task method might have a slight advantage, but this should not be overemphasized as this might be a result of the differences in the force calculation code. 

For the bulk system, the cell task method obtains a parallel efficiency of 83.6\,\% when using 12 threads. 
One might ask whether for a system with more than one million atoms this value should not be higher since for a system of this size the fraction of time spent in serial code can be expected to be negligible. While some overhead is certainly caused by the dynamic scheduling of the tasks the principal factor limiting parallel efficiency is memory access speed. A short test on a system with a lower clock frequency but faster memory resulted in a higher parallel efficiency of 90.2\,\% for 12 threads. In addition to this, the fact that the spatial decomposition program achieves similar speedups indicates  that the parallel efficiency is limited by hardware constraints rather than the algorithms.  

\subsection{Copper nanoparticle}
\begin{figure}
\includegraphics[width=7.2cm]{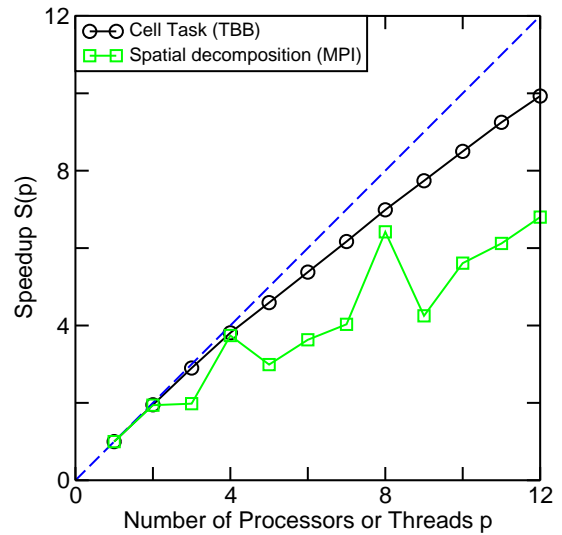}
\caption{(Color online) Parallel speedup factors in simulations of  a spherical copper nanoparticle as a function of the number of threads or processors for the cell task method (black circles) and spatial decomposition (green squares).The dashed blue line represents the ideal speedup of one per processor.}\label{FigSphere} 
\end{figure}
The spherical nanoparticle system is inhomogeneous in the sense that the particle fills only a part of the volume of the simulation box leaving the rest empty. For the spatial decomposition method this configuration is challenging since there is no simple way to divide the simulation box into an arbitrary number of domains with equal shapes so that each domain contains a similar number of particles. Exceptions are the cases of two, four and eight processors where equal partitions can be obtained by cutting the sphere repeatedly into halves. This is confirmed by the speedup factors obtained on the multi-core machine for this system (see Fig.~\ref{FigSphere}).  Spatial decomposition in this system is efficient only for two, four and eight processors. In all other cases the parallel efficiency is significantly lower in the range of 50\,\% - 60\,\%. In contrast to this, the cell task method yields efficiencies above 80\,\% for all numbers of threads. The parallel efficiency of the cell task method for this system is 82.8\,\% when using 12 threads, very similar to the value obtained for the bulk system.

%
%
\subsection{Partially sintered nanocrystalline copper}
\begin{figure}
\includegraphics[width=7.2cm]{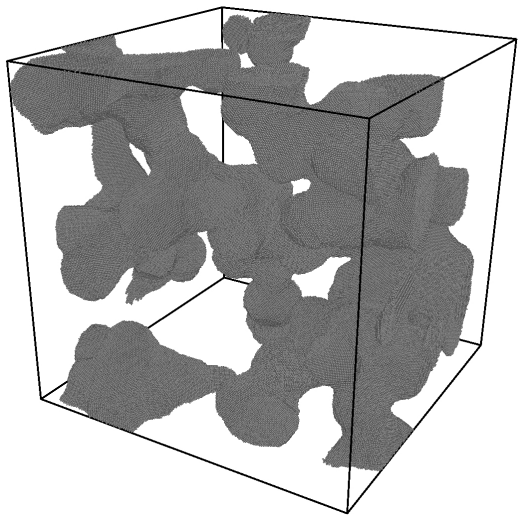}
\caption{Configuration of the partially sintered nanocrystalline copper system containing 1,992,220 atoms. }
\label{FigCuboCfg}
\end{figure}

The  partially sintered nanocrystalline copper system is the most complex and inhomogenous of the three benchmark systems (see Fig.~\ref{FigCuboCfg}). The configuration is an intermediate result from a simulation of the coalescence of an ensemble of copper nanoparticles. As can be seen from Fig.~\ref{FigCuboCfg} the system is very inhomogeneous with an irregular distribution of the atoms and a large amount of empty space. The work presented in this article was motivated by systems like this.

\begin{figure}
\includegraphics[width=7.2cm]{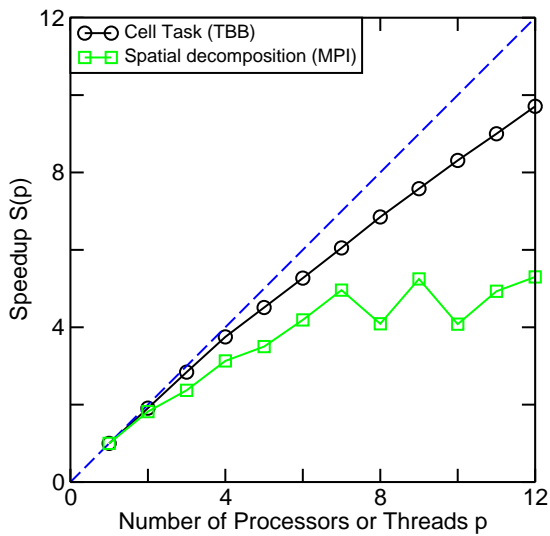}
\caption{(Color online) Parallel speedup factors in simulations of partially sintered nanocrystalline copper as a function of the number of threads or processors for the cell task  method (black circles) and spatial decomposition (green squares).The dashed blue line represents the ideal speedup of one per processor.}\label{FigCubo} %
\end{figure}

Figure~\ref{FigCubo} shows the speedup factors obtained for the nanocrystalline configuration on the multi-core machine. It is clear from this figure that spatial decomposition is not very effective for this kind of system. While the speedup factors increase more or less monotonously, they are far below the ideal value of one per processor. The reason for this is clear from Fig.~\ref{FigCuboCfg}. Only by great chance could  a regular subdivision of the system into domains result in a balanced workload. 

The cell task method, on the other hand, has no particular problems with the porous nature of this system. The speedups shown in Fig.~\ref{FigCubo} for the cell task method are very similar to those obtained for the two previous benchmark systems. With 80.9\,\% the parallel efficiency of the cell task method for this system when using 12 threads is slightly lower than for the previous two systems. This can probably be explained by the large surface of the configuration which reduces the average number of neighbors per particle significantly. The lower number of neighbors in turn affects the memory access pattern and might reduce cache efficiency.

%
%
\subsection{Pair of sintered copper nanoparticles}\label{SecSys08}
\begin{figure}
\includegraphics[width=7.2cm]{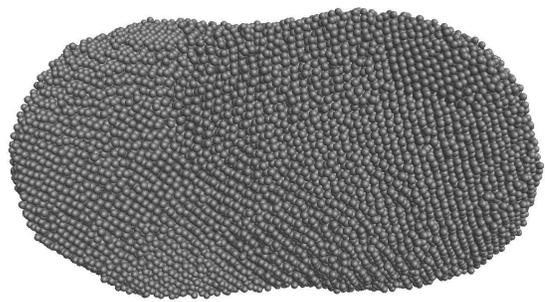}
\caption{Configuration of two sintered copper nanoparticles (57,482 atoms). }%
\label{FigSys08Cfg}%
\end{figure}
In order to test the efficiency of the cell task method for smaller configurations, a system consisting of two sintered copper nanoparticles with only 57,482 atoms was used. Fig.~\ref{FigSys08Cfg} shows the dumbbell-like shape of the system, which constitutes a similar challenge to spatial decomposition as the spherical nanoparticle.
 
\begin{figure}
\includegraphics[width=7.2cm]{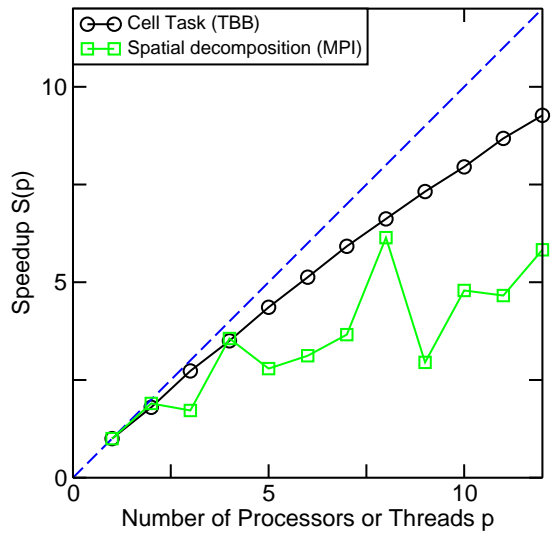}
\caption{(Color online) Parallel speedup factors in simulations of two sintered copper nanoparticles as a function of the number of threads or processors for the cell task  method (black circles) and spatial decomposition (green squares).The dashed blue line represents the ideal speedup of one per processor.}%
\label{FigSys08}%
\end{figure}
On the multi-core machine, the behavior of the speedup factor in this benchmark is similar to the nanoparticle system (see Fig.~\ref{FigSys08}). The speedup of the cell task method increases continuously with the number of threads. The speedups for the spatial decomposition method are generally inferior to those obtained by the cell task method except for $p = $ 2, 4 and 8 where the symmetry of the configuration allows an efficient decomposition. 

The main difference between the speedups of the cell task method shown in Fig.~\ref{FigSys08} and the other benchmark systems is a slight overall reduction of the cell task method's parallel efficiency. When using 12 threads, the efficiency obtained for this system is only 77.3\,\%. The reason for this reduction is most likely the growing impact of serial code.

%
%
\subsection{Simulations on Xeon Phi coprocessor}
The benchmark results presented in the preceding sections demonstrate the efficiency of the cell task method on typical multi-core machines. In this section the efficiency of the method on a many-core system is studied.

The Intel Xeon Phi coprocessors 5110P integrates 60 compute cores on a single chip and provides a large memory bandwidth. Each core supports four hardware threads and has 512-bit vector registers that allow the simultaneous execution of eight double-precision floating-point operations. Under normal conditions, optimal performance can be achieved only if at least two threads are running on each core.  For a detailed discussion of these devices the reader is referred to Ref.~\cite{JeffersReinders}. 

\begin{figure}
\includegraphics[width=7.2cm]{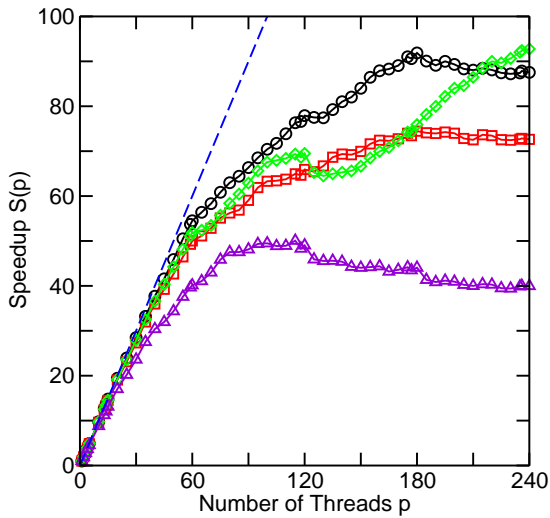}
\caption{(Color online) Parallel speedup of the cell task method on a Xeon Phi coprocessor. Benchmark systems are bulk Cu (black circles), Cu nanoparticle (green diamonds), partially sintered nanocrystalline copper (red squares), and a pair of Cu nanoparticles (purple triangles). The dashed blue line indicates the ideal speedup of one per thread.}
\label{FigMic}
\end{figure}
Figure~\ref{FigMic} shows the parallel speedups obtained with the four benchmark systems used in this work on the Xeon Phi coprocessor using up to 240 threads. Interestingly, the behavior of all four systems is different.  

Until $p=60$ threads, the speedups for the three large systems stay close to the line of ideal speedup. At $p=60$, the speedup of the three systems range from $50.0 - 54.5$ (cf. Table~\ref{TabMic}), corresponding to parallel efficiencies between 83.4\,\% and 90.8\,\%.  At this point, the three curves exhibit a sharp bend but the speedup continues to rise at a lower rate until $p=120$. This is due to the fact that now more than one thread is running per core. Although the Xeon Phi microarchitecture requires more than one thread per core for optimal performance, it is not guaranteed that the performance will be doubled by a second thread. The threads share after all some of the physical units of the core like level 1 and 2 caches. A likely reason for the slower increase of the speedup for $p$ in the range from 60 to 120 is an increased amount of cache misses. 

Beyond $p=120$ the speedup curves of the three large benchmark systems show a more or less pronounced dip (hardly noticeable for the porous nanocrystalline system and most visible in case of the nanoparticle) after which the speedup continues to rise. The reason for this dip is not yet understood. The bulk system and the porous nanocrystalline system reach their maximum speedup at $p=180$ after which the speedup decreases for both systems. In contrast to this, the nanoparticle simulation increases its speedup until $p=240$. 

Compared to the larger benchmark systems, simulations of the sintered nanoparticles achieve significantly lower speedups on the Xeon Phi. The speedup curve for this system deviates early on from the ideal behavior and it reaches a plateau around $p=100$. For larger values of $p$ the curve decreases with clear steps at $p=120$ and $p=180$.
The lower performance of the Xeon Phi in this benchmark can be explained by the system size. The small volume of the configuration does not allow for more than about 100 tasks to be run simultaneously. Adding more threads can therefore not accelerate the computations. The decrease of the efficiency for larger values of $p$ is probably caused by increased management overhead when the internal task queues run empty. A similar effect might be the reason for the efficiency decrease in case of the bulk system beyond $p=180$.

\begin{table}
\caption{
Speedup of benchmarks systems on the Xeon Phi coprocessor: Number of threads $p_\mathrm{max}$ for which the maximum speedup was obtained together with the corresponding speedup $S(p_\mathrm{max})$, and the speedup for 60 threads $S(60)$ and maximum speedup $S_\mathrm{ref}(p_\mathrm{max})$ with respect to the 12-core test machine.}\label{TabMic}
\begin{tabular}{lcccc}
\hline\hline
System & $p_\mathrm{max}$ & $S(p_\mathrm{max})$ & S(60) & $S_\mathrm{ref}(p_\mathrm{max})$\\
\hline
Bulk 					& 180 & 91.9 & 54.5 & 11.8 \\
Nanoparticle 			& 240 & 92.8 & 50.0 & 13.4 \\
Porous nanocrystalline	& 180 & 74.5 & 51.9 & \phantom{0}9.8 \\
Sintered nanoparticles 	& 115 & 50.1 & 40.1 & \phantom{0}7.3 \\
\hline\hline%
\end{tabular}
\end{table}
Table~\ref{TabMic} summarizes the maximum speedups obtained for the four benchmark systems on the Xeon Phi coprocessor. The best speedups above 90 are achieved by the bulk system and the nanoparticle. The porous nanocrystalline system reaches a maximum speedup of 75, whereas the small system of two sintered nanoparticles is limited to $S(p_\mathrm{max})=51$. 

The table also gives the speedup of the systems on the coprocessor with respect to single thread execution on the 12 core machine. These values are not overly impressive. One has to keep in mind, however, that the individual cores of the coprocessor are considerably less powerful than a modern general purpose CPU. As pointed out in Ref.~\cite{JeffersReinders}, the key to a strong performance of the Xeon Phi coprocessor is a highly parallel algorithm combined  with vectorization. In this work only the first part has been addressed.    

%
%
\section{Summary and Conclusions}
This article describes the design of a parallel algorithm for MD simulations with short-ranged forces on single node multi- and many-core systems. The aim of the cell task algorithm is to provide an efficient parallelization method for systems where  spatial decomposition is not effective. Examples for such problematic cases are large inhomogeneous systems like nanostructured materials or nanodevices. 

The cell task method makes use of the  linked-cell technique to subdivide the force calculation into small tasks. The tasks are then executed by a team of threads according to a dependent task schedule. This schedule is an important part of the algorithm. It avoids the situation that a particle is accessed by two threads simultaneously. This effectively eliminates the need for synchronization constructs in the force calculation, which makes the algorithm very efficient.    

Benchmark simulations on a 12-core system show clearly that for inhomogeneous configurations like nanoparticles or porous systems the cell task method performs significantly better than spatial decomposition. For a homogeneous crystalline bulk system both methods achieve comparable speedups.  An added advantage of the cell task method is its consistent performance. The speedup factor of this method increases continuously with the number of threads in a very similar manner for all  types of systems. It relieves the user from technical considerations like which spatial decomposition strategy works best for a given system.

The benchmarks on the multi-core system reveal two factors that limit the parallel efficiency of the cell task method: memory access speed and small system size. Neither of these factors is specific to the cell task method. No parallelization strategy will work if the memory system cannot deliver the data fast enough, and every MD program contains inevitably some serial code that limits the parallel efficiency. Note, however, that in the current implementation of the cell task method the construction of the task schedule increases the fraction of time spent in serial code slightly as the wave algorithm has not been parallelized yet.

Execution of the benchmark simulations on a 60 core Intel Xeon Phi coprocessor shows that the cell task method scales  to large numbers of threads. For large configurations the speedup increases linearly and close to the ideal line until $p=60$. 
Beyond 60 threads, the speedups increase at a lower rate since the number of threads per core is now greater than one. Running more than one thread per core is almost always beneficial on the Xeon Phi since it hides latencies of the microarchitecture~\cite{JeffersReinders}. Table~\ref{TabMic} shows that the three larger benchmark systems reach their maximum speedups on the Xeon Phi at three or four threads per core. These maximum speedups $S(p_\mathrm{max})$ range  from 75 to 93 and they are 44\,\% to 86\,\% higher than the speedups $S(60)$ obtained at one thread per core. 

The smallest benchmark system does not perform optimally on the Xeon Phi coprocessor. The reason for this is that the configuration is not large enough to keep two or more threads per core busy. The dependent task schedule limits the number of tasks that can run simultaneously to at most 1/27 of the total number of tasks. In order to keep all threads running, the number of tasks must therefore be at least 27 times the number of threads. In practice optimal performance might require an even substantially larger number of tasks per thread. More work will be necessary to study the performance limits of the algorithm for smaller systems.

All benchmarks  shown in this work used the copper potential by Cleri and Rosato~\cite{Cleri:93a}. This choice might have limited the parallel efficiency due to the low computational complexity of the potential which emphasizes memory access issues. It would be interesting to see the performance of the cell task algorithm with a computationally more demanding potential like, for example, one of the EAM potentials in Ref.~\cite{Mendelev:03a}. With such a potential higher parallel efficiencies closer to the ideal limit can be expected since the influence of memory access speed and serial code execution would be diminished. 

Work is currently in progress to let the program take advantage of the vector capabilities of the Xeon Phi architecture. In addition to this it is planned to extend the method into a hybrid scheme for distributed systems that combines the cell task algorithm for the parallelization on the compute nodes with spatial decomposition to share the work between multiple nodes.

\begin{acknowledgments}
This work has been supported financially by Laurentian University and  the Natural Sciences and Engineering Research Council of Canada (NSERC). Generous allocation of computer time on the facilities of the  Shared Hierarchical Academic Research Network (SHARCNET)~\cite{SHARCNET} and Compute/Calcul Canada is gratefully acknowledged.
\end{acknowledgments}

\appendix*
\section{Generation of Task Waves}
The wave algorithm generates the dependent task schedule for the cell task method from a series of sets of non-overlapping tasks (\textit{waves}). This appendix gives details about the construction of the waves.

The indices of the cells belonging to one wave are generated from the Cartesian product of three sets of integers (one for each dimension). To give a two-dimensional example, the wave shown in Fig.~\ref{FigWave} could be constructed from the product $\{2,5\} \times \{2,5\}$ [we assume here and in the following that the cell in the lower left corner has the indices (1,1)]. The product $\{3,6\} \times \{2,5\}$ would generate a similar wave with the tasks shifted one cell to the right. Note that not all index sets can be used to generate a wave. In order to obtain non-overlapping task, the differences between all indices in a set must be at least three (subject to the boundary condition of that dimension; for a dimension with $D$ cells and periodic boundary conditions, the difference between index 1 and $D$ is 1). The product  $\{2,4\} \times \{2,5\}$, for example, does not give a valid wave.

The number of index sets necessary to cover all cells in one dimension depends on two factors. If the number of cells $D$ in that dimension is a multiple of three or if periodic boundary conditions do not apply in that direction, three index sets are enough. These three sets start with the numbers 1, 2, and 3, respectively, and advance in steps of three to the end of the system. For example, all cells of a non-periodic system with $D=14$ are covered by the following three sets: $\{1,4,7,10,13\}$, $\{2,5,8,11,14\}$, and $\{3,6,9,12\}$. This coverage of the cells in one dimension is shown by the following figure where the numbers indicate to which wave the cell belongs.
\begin{displaymath}\begin{array}{|c|c|c|c|c|c|c|c|c|c|c|c|c|c|}
\hline
1 & 2 & 3 & 1 & 2 & 3 & 1 & 2 & 3 & 1 & 2 & 3 & 1 & 2\\
\hline
\end{array}\end{displaymath}
It is easy to see that the cells belonging to the same wave are separated by at least two cells not belonging to that wave. 
 
Things are more complicated for directions where periodic boundary conditions apply and where $D$ is not a multiple of three. In this case a fourth index set is required. The index sets for such a case can be obtained by advancing continuously in steps of three, wrapping around at the boundary $D$. For a periodic system with $D=14$, the first set is then $\{1,4,7,10\}$. Unlike the non-periodic case, $13$ cannot be included in this set as that cell overlaps with the first. By advancing in steps of three we obtain the second set $\{13, 2, 5,8\}$. Advancing further in steps of three we obtain for the third and fourth sets  $\{11, 14, 3,6\}$ and $\{9,12\}$. The coverage of the cells in this case is thus 
\begin{displaymath}\begin{array}{|c|c|c|c|c|c|c|c|c|c|c|c|c|c|}
\hline
1 & 2 & 3 & 1 & 2 & 3 & 1 & 2 & 4 & 1 & 3 & 4 & 2 & 3\\
\hline
\end{array}\end{displaymath}

A complete set of waves in more than one dimension can be generated by nested loops where each loop generates the index sets for one dimension. In order to keep the number of dependencies between tasks low, it is best if from one wave to the next, the task pattern is shifted only in one direction. This is achieved if the inner loops are not reset when one of the outer loops advances. Instead, the inner loops repeat their last pattern, possibly filling it to the maximum number of cells. This is explained by the next figure, which shows the first six waves in a periodic system with $14\times 6$ cells:
\begin{displaymath}\begin{array}{|c|c|c|c|c|c|c|c|c|c|c|c|c|c|}
\hline
 &  &  &  &  &  &  &  &  &  &  &  &  & \\
\hline
5 &   & 6 & 5 &  &  & 6 &  & 5 &  & 6 & 5 &  & 6\\
\hline
1 & 2 & 3 & 1 & 2 & 3 & 1 & 2 & 4 & 1 & 3 & 4 & 2 & 3\\
\hline
 &  &  &  &  &  &  &  &  &  &  &  &  & \\
\hline
5 &   & 6 & 5 &  &  & 6 &  & 5 &  & 6 & 5 &  & 6\\
\hline
1 & 2 & 3 & 1 & 2 & 3 & 1 & 2 & 4 & 1 & 3 & 4 & 2 & 3\\
\hline
\end{array}\end{displaymath}

The first four waves use the same index set $\{1,4\}$ for the $y$ direction and thus essentially repeat the one-dimensional example given above. When going from the fourth to the fifth wave, the index set for the $x$ direction does not return to the set of the first wave. Instead the set reuses the indices from the fourth wave $\{9, 12\}$ and completes the set by advancing in steps of three which leads to leads to the set $\{9, 12, 1, 4\}$. The sixth set is then generated by further advances in steps of three as described above.

Since each dimension is covered by three or four sets of indices, the total number of waves for a system with $d$ dimensions is given by a product of $d$ factors where each factor is 3 or 4 depending on the dimensions of the cell grid and the boundary conditions. In three dimensions the total number of waves is thus  27, 36, 48, or 64.  

%
%
%

\end{document}